\documentclass[prd,twocolumn,floatfix,amsmath,nofootinbib,amssymb,superscriptaddress]{revtex4}
\usepackage{graphicx,color,dcolumn,booktabs,bm}
\usepackage{longtable,lscape}
\usepackage{pdfpages}
\usepackage{txfonts}
\usepackage{overpic}
\usepackage{amssymb}
\usepackage{makecell}
\usepackage{indentfirst}
\usepackage{feynmf}
\usepackage{slashed}
\usepackage[utf8]{inputenc}
\usepackage{cases}
\usepackage{color}
\usepackage{multirow}
\usepackage{threeparttable}
\usepackage{epstopdf}
\usepackage{enumerate}
\usepackage{subfigure}
\usepackage{diagbox}
\usepackage{graphicx,color,dcolumn,booktabs,bm}
\usepackage{mathrsfs}
\usepackage{cancel}
\usepackage{float}
\usepackage[misc]{ifsym}
\usepackage[colorlinks,
citecolor=blue,
anchorcolor=red,
menucolor=red,
linkcolor=red,
filecolor=red,
runcolor=red,
urlcolor=blue,
frenchlinks=true]{hyperref}
\usepackage{array}
\usepackage{booktabs}
\usepackage{amsmath}
\usepackage{slashed}
\usepackage{tikz}
\usetikzlibrary{
	arrows,
	arrows.meta,
	decorations.pathmorphing
}
\begin{document}
\title{On compositeness of  $D_{s0}^*(2317)$ and its decay to $D_s\pi^0$}

\author{Yu-Hui Zhou}
\affiliation{School of Physics, Southeast University, Nanjing 211189,
P.~R.~China}
\author{Xiu-Li Gao}
\affiliation{School of Physics, Southeast University, Nanjing 211189,
	P.~R.~China}

\author{Bin Wu}

\affiliation{School of Physics, Southeast University, Nanjing 211189,
P.~R.~China}
\author{Jun-Xi Cui}
\affiliation{School of Physics, Southeast University, Nanjing 211189,
P.~R.~China}
\author{Zhi-Yong Zhou}
\email[Corresponding Author:]{zhouzhy@seu.edu.cn}
\affiliation{School of Physics, Southeast University, Nanjing 211189,
P.~R.~China}

\date{\today}

\begin{abstract}
We demonstrate that  the  $D_{s0}^{*}(2317)$ can be described as a bound-state pole arising from the coupling between a discrete  $c\bar{s}(1^3P_0)$ state and the  $DK$ continuum state in the Lee-Friedrichs model. The elementariness and compositeness of the $D_{s0}^{*}(2317)$  are determined  to be about $Z:X\approx 51.1   \%:48.9\%$, indicating a nearly equal admixture of compact quark-model state and hadronic molecular components. The decay width of $D_{s0}^{*}(2317)\rightarrow D_s^{+}\pi^0$ is evaluated within the quark rearrangement framework. The transition  $c\bar{s}(1^3P_0)\rightarrow D_s\pi^0$ proceeds via the OZI-allowed $c\bar{s}(1^3P_0)-D_s\eta$ coupling followed by $\eta-\pi^0$ mixing, which serves as the primary source of isospin violation in this channel. The coupling between $DK$ and $D_s\pi^0$ is computed in a quark rearrangement model, where the isospin violation effect originates from the mass difference between the  $D^0K^+$ and $D^+K^0$ thresholds. The parameters of the scheme shares the same ones with those of the underlying potential model and only the vacuum production strength is adjusted  to reproduce the physical $D_{s0}^{*}(2317)$ mass. 
We examine the sensitivity of our results to the only free parameter, the pair-creation strength $\gamma$, and discuss the systematic uncertainties associated with the truncation of coupled channels and the underlying phenomenological potential model.
This calculation may shed more insight to the nature of exotic $D_{s0}^{*}(2317)$ state and its isospin-breaking decay properties.
\end{abstract}

\maketitle
\section{Introduction}
The $D_{s0}^{*}(2317)$ was first observed in the $D_s^+\pi^0$ invariant mass spectrum by the   BABAR  collaboration~\cite{BaBar:2003oey} and soon confirmed by the CLEO~\cite{CLEO:2003ggt} and Belle~\cite{Belle:2003guh}. It has attracted  considerable attention due to its unusual properties:
it is a narrow resonance lying below the $DK$ threshold, with a precisely determined mass of $2317.8 \pm 0.5$ MeV and  quantum numbers $J^P=0^{+}$.
Recently, the Belle~II Collaboration updated the ratio $R$ of the radiative decay 
$D_{s0}^{*}(2317)\rightarrow D_{s}^{*}\gamma$
to the isospin-violating strong decay $D_{s0}^{*}\rightarrow D_{s}^{+}\pi^{0}$ ~\cite{Belle-II:2025dzk}. This new result deviates noticeably from the value previously quoted  by the PDG~\cite{ParticleDataGroup:2024cfk}, renewing interest in its internal structure.

From the perspective of conventional quark models, the $D_{s0}^{*}(2317)$  presents a significant challenge. For example, the Godfrey–Isgur~(GI) model~\cite{Godfrey:1985xj} predicts the lowest $c\bar{s}(1 ^{3}P_{0})$ state at around 2.480 GeV, about 100 MeV above the observed mass of the $D_{s0}^{*}(2317)$. Similarly, the relativistic potential model proposed by Ebert, Faustov and Galkin  also presents that the $c\bar{s}(1 ^{3}P_{0})$ is located at about 2.509 GeV~\cite{Ebert:2009ua}. A further distinctive feature of this state is its mass lying in close proximity below the open-flavor $DK$ threshold, where two-body hadronic dynamics can significantly influence  spectroscopic properties. In recent years, many exotic candidates  near two-body  thresholds have been observed experimentally, such as $X(3872)$, $Z_{c}(3900)$, $f_{0}(980)$, $a_{0}(980)$~\cite{Acosta:2003zx,BESIII:2013ris,Belle:2013yex,Hyams:1973zf,Astier:1967zz} and so on. The common feature of these states strongly suggests that exotic configurations beyond the conventional quark model may play an essential role in their structures.  These observations have stimulated extensive experimental and theoretical efforts aimed at elucidating the internal structures of such states.

Unlike many  other exotic candidates, the $D_{s0}^{*}(2317)$ decays primarily via the isospin-violating $D_{s}\pi^{0}$ mode, resulting in an extremely narrow total width. This makes a direct experimental measurement of the width particularly difficult. Currently, only  an upper limit of its width, $\Gamma<3.8$ MeV, is available. Consequently, understanding the internal structure of this exotic state therefore provides an important testing ground for models of strong interaction dynamics in the heavy–light sector.
Various theoretical scenarios have been proposed to study the structure of this state, including a compact $c\bar{s}$ meson~\cite{Wang:2006bs,Han:2023wqq,Dai:2003yg,Bali:2003jv,Song:2015nia,Zhong:2008kd,Wei:2005ag,Chen:2009zt,Hwang:2004cd,Cheng:2014bca,Luo:2021dvj,Alhakami:2016zqx,Deandrea:2003gb}, a $DK$ hadronic molecule~\cite{Barnes:2003dj,Guo:2008gp,Liu:2012zya,Alexandrou:2019tmk,Faessler:2007gv,Fu:2025lfo,Chen:2004dy,Yeo:2024chk,Zhou:2025rpb,Kolomeitsev:2003ac,Szczepaniak:2003vy,Xie:2010zza,Cleven:2010aw,Du:2017ttu,Guo:2018tjx,Su:2025aiz}, a tetraquark configuration~\cite{Alexandrou:2019tmk,Chen:2004dy,Wang:2006uba,Nielsen:2005ia,Nielsen:2005zr}, or a mixed state~\cite{vanBeveren:2006st,Yang:2021tvc,Simonov:2004ar,Dai:2006uz,Dong:2017gaw,Ni:2023lvx,Zhou:2020moj} containing both quark–antiquark and meson–meson components. 
%Numerous studies indicate that, the decay width of   $D_{s0}^{*}(2317)\rightarrow D_s^+\pi^0$ when treated as a pure  $c\bar{s}$ meson is significantly smaller than as a $DK$ molecular picture.
Theoretical studies indicate that the decay width of 
$D_{s0}^{*}(2317)\rightarrow D_s^+\pi^0$ 
is significantly smaller in a pure $c\bar{s}$ picture than in the $DK$ molecular scenario.
It can be understood intuitively that, if the state is a compact 
$c\bar{s}$ meson, the decay can only proceed through 
$c\bar{s}\rightarrow D_s\eta$ followed by  $\eta-\pi^{0}$
mixing, whereas in the $DK$ molecular picture the decay is enhanced by the S-wave coupling to the $DK$ channel. 
Combining the lattice QCD with the unitarized chiral perturbation theory~\cite{Liu:2012zya}, the  $DK$ attractive  interaction generates a pole located at 2317 MeV, and results in an isospin-breaking decay width to $D_s^+\pi^0$ about 133 keV.
By comparison,
Ref.~\cite{Yang:2021tvc} investigated the coupled-channel effects of $D_{s0}^{*}(2317)$ using a quark-model framework constrained by lattice QCD inputs, and found that the state consists of about 32\% bare  $c\bar{s}$ core and 68\% $DK$ component. 
The study also indicates that $DK$ interaction alone  is insufficient to reproduce the $D_{s0}^{*}(2317)$.
Namely, $D_{s0}^{*}(2317)$ is favored to be interpreted as a mixed state rather than a pure $DK$ molecular state in Ref.~\cite{Yang:2021tvc}.
Similar conclusions can be drawn in the unquenched quark model, where
the strong coupling between the $c\bar{s}$ core and the $DK$ naturally accounts for the sizable mass shift of the $D_{s0}^{*}(2317)$, 
$\Delta M \simeq 100~\text{MeV}$, between the physical and bare $c\bar{s}$ states~\cite{Ni:2023lvx,Ni:2021pce}.

Despite extensive theoretical efforts   devoted to decoding the $D_{s0}^{*}(2317)$, 
there is still no consensus both on its internal composition and the value of its width. 
Based on a preliminary survey, the predicted values of decay width span from a few keV up to over 200 keV~\cite{Faessler:2007gv,Fu:2021wde,Fajfer:2015zma,Yue:2025wcl,Wu:2014era,Liu:2006jx,Nielsen:2005zr,Colangelo:2003vg,Wei:2005ag,Ishida:2003gu,Zhou:2025rpb,Guo:2008gp,Cleven:2014oka,Mehen:2004uj,Lu:2006ry,Bardeen:2003kt,Godfrey:2003kg,Achasov:2025hkv},
with a number of them displaying some degree of model dependence.

Motivated by the current incomplete understanding of this state, we investigate the  $D_{s0}^{*}(2317)$  within the extended Lee-Friedrichs (LF) model, with a focus on its compositeness and strong decay properties.
Within this framework, once a pole appears on the physical Riemann sheet, the compositeness of the state, namely the related contributions from the bare state $c\bar{s}$ and the coupled continuous $DK$ channels, can be determined in a rigorous and model-consistent manner.
The strong decay width for $c\bar{s}\rightarrow D_{s}^{+}\pi^{0}$ 
is then evaluated by quark pair creation (QPC) model through the OZI-allowed process $c\bar{s}\rightarrow D_{s}\eta$, following a $\eta-\pi^{0}$ mixing.
The transition amplitude of $DK\rightarrow D_{s}\pi^{0}$ can be computed by the quark rearrangement model.
Naturally, the total decay width  is subsequently obtained by combining these contributions coherently. 
It is worth emphasizing that no additional free parameters are introduced beyond the vacuum pair creation strength $\gamma$ , which makes our prediction particularly constrained and free from superfluous assumptions. 

This article is organized as follows. 
The theoretical framework is introduced  in Sec.~\ref{sec.theory}, which consists of the extended Lee-Friedrichs model, the QPC model and the quark rearrangement model.
The detailed calculation procedure and the main results are presented in Sec.~\ref{result}.
Finally, a brief  summary and the conclusions are given.

\section{Theoretical framework}
\label{sec.theory}
\subsection{Wave function of $D_{s0}^*(2317)$ in the extended Lee-Friedrichs scheme}

In the extended Lee-Friedrichs scheme~\cite{Xiao:2016dsx,Xiao:2016mon,Xiao:2016wbs,Xiao:2023lpv} the full Hamiltonian $H$ can be expressed as
\begin{equation}
	\begin{aligned}
		\label{fullHamiltonian}
		H=&m_0 |0\rangle\langle 0| +\sum_{n,S,L}
		\int_{M_{n}}^\infty \mathrm \, \mathrm{d}E\, E |E,[n]_{SL}\rangle \langle E,[n]_{SL}|\\
		&+\sum_{n,S,L}\int_{M_{n}}^\infty \, \mathrm{d}E  f_n^{SL}(E)|0\rangle\langle
		E,[n]_{SL}|+h.c.,\\
	\end{aligned}
\end{equation}
where $m_0$ denotes the bare mass of the discrete state $|0\rangle$,  $[n]_{SL}$ the
$n$-th species continuum state with the total spin $S$ and the angular momentum $L$, $M_{n}$ the energy threshold of $n$-th
continuum state, $S$ and $L$ the total spin and the angular momentum
of the continuum states, $f_n^{SL}$ the coupling functions between the
discrete state and the continuum state. 
The eigenvalue problem of the full Hamiltonian in
Eq.~(\ref{fullHamiltonian}) is exactly solvable in the rigged Hilbert space.
 The eigenlvalues of  bound states, virtual states or resonant
states in the scattering amplitude  can be obtained by finding zeros of the inverse of resolvent function  $\eta(z)=0$ on the complex energy plane where
\begin{align}
	\eta(z)&=z-m_0-\Pi(z)\nonumber\\
    &=z-m_0-\sum_{n,S,L}\int_{M_{n}}^\infty\frac{|f^{SL}_n(E)|^2}{z-E
	}\mathrm{d}E,
	\label{resolvent}
\end{align}
and the wave functions of these states can also be explicitly written down~\cite{Civitarese200441,Xiao:2016dsx}.
For the general case of one discrete state coupled to multiple
continuum channels, the corresponding coupled-channel scattering matrix
can be expressed in matrix form as~\cite{Xiao:2023lpv}
\begin{equation}
	S=I+2\pi iT
	=I+2\pi iF^\dagger\eta^{-1}F,
	\label{Smatrix}
\end{equation}
where $F$ denotes the coupling matrix between the bare discrete states
and the continuum channels. Since the Lee--Friedrichs formalism is
manifestly unitary, the scattering matrix defined by
Eq.~(\ref{Smatrix}) automatically satisfies the coupled-channel
unitarity condition.

In principle, the discrete state can couple to all continuum channels with allowed quantum numbers.
However, one usually has to restrict the calculation to a finite number of the most relevant OZI-allowed channels.
For the $D_{s0}^{*}(2317)$, the $DK$ channels are expected to be dominant, since the state lies immediately below the $DK$ thresholds, leading to an enhanced near-threshold contribution through the energy denominator in the wave function.
Contributions from higher-lying channels are expected to be relatively suppressed.
Moreover, for channels involving $s\bar{s}$ pair creation, such as $D_s\eta$, an additional suppression is generally expected relative to channels involving $u\bar{u}$ or $d\bar{d}$ pair creation in phenomenological models at the quark level.
We therefore retain only the dominant continuum channels $D^0K^+$ and $D^+K^0$,  in this framework.
The physical state $D_{s0}^*(2317)$ could be represented as a bound-state pole in this picture, which is mainly determined by the bare state and the relevant continuum states.
The bare mass is chosen to be the predicted $1^3P_0$ $c\bar{s}$ state of the potential model~\cite{Godfrey:1985xj} and the coupling functions are computed via QPC model~\cite{Micu:1968mk,Blundell:1995ev}. The wave function of $D_{s0}^*(2317)$ could be explicitly written down as the superposition of the discrete bare $c\bar{s}(1^3P_0)$ and the continuous $D^0K^++D^+K^0$ state as
\begin{equation}
		\label{eq:Ds2317-wave-funtion}
		\begin{aligned}
	  |D_{s0}^*(2317)^+\rangle=&N_B\bigg(\left|c\bar s(1^3P_0)\right\rangle+\int_{M_{0+}}^\infty
	  \frac{f_{0+}(E)} {z_{X}-E}\left|E,[D^{0}K^{+}]\right\rangle \, \mathrm{d}E\\
	  +&\int_{M_{+0}}^\infty
	  \frac{f_{+0}(E)}
	  {z_X-E}\left|E,[D^+K^0]\right\rangle \, \mathrm{d}E \bigg)+\cdots,
	  \end{aligned}
\end{equation}
where  $N_B$ is the normalization factor and the quantum number $S=0$, $L=0$ is implicit in this case. $z_X$ is the pole position which is related to the physical mass of the $D_{s0}^*(2317)$. The $\cdots$ represents other minor contributions from higher-lying coupled channel and the open weakly interacting channels including $D_s\pi^0$.

\subsection{The transition amplitude of the isospin violating decay channel}

Since the wave function of $D_{s0}^*(2317)$ is only represented by the predominant components, its decay to $D_s\pi^0$ is treated in a perturbative manner to the leading order. 
Define the decay process $A\rightarrow BC$ and its $S$ matrix as 
\begin{equation}
	S_{fi}=\delta_{fi}-2\pi i\delta(E_f-E_i)\langle BC|H_I|A\rangle
\end{equation}
and the isospin-breaking decay amplitude of $D_{s0}^*(2317)$ to the partial $L'$-wave $D_s\pi^0$ channel with total spin $S'$ is defined as
\begin{equation}
	\begin{aligned}
    &M_{S^{'}L^{'}}\left(D_{s0}^{*}(2317)\rightarrow D_{s}\pi^{0}\right)\\
	=&\left\langle [D_s^+\pi^0]_{S'L'}|H_I|D_{s0}^*(2317)^+\right\rangle\\
	=&N_B\Bigg(\left\langle [D_s^+\pi^0]_{S'L'}\left|H_I\right|c\bar s(1^3P_0)\right\rangle\\
	+&\int_{M_{0+}}^\infty
	\sum_{SL}\frac{f_{0+}^{SL}(E)} {z_X-E}\left\langle [D_s^+\pi^0]_{S'L'}\left|H_I\right|[D^0K^+]_{SL}\right\rangle\mathrm{d}E\\
	+&\int_{M_{+0}}^\infty
	\sum_{SL}\frac{f_{+0}^{SL}(E)}
	{z_X-E}\left\langle [D_s^+\pi^0]_{S'L'}\left|H_I\right|[D^+K^0]_{SL}\right\rangle\mathrm{d}E\Bigg).
	\end{aligned}
	\label{eq:Ds2317-decay}
\end{equation}
One must also notice here $S=L=S'=L'=0$ in this case and the summation is only listed for completeness of the formula.
If the isospin is conserved, the interaction matrix elements
$\langle [D_s^+\pi^0]_{S'L'}|H_I|c\bar s(1^3P_0)\rangle$  vanishes and the two later terms of Eq.~(\ref{eq:Ds2317-decay})
 cancel each other.  In this paper, the isospin  violation effect for the $\left\langle [D_s^+\pi^0]_{S'L'}\left|H_I\right|c\bar s(1^3P_0)\right\rangle$ term is assumed to be contributed through the intermediate OZI-allowed coupling of $[c\bar s(1^3P_0)]-D_s\eta$ process following by the $\pi^{0}-\eta$  mixing mechanism. The contributions of $DK$ continuum contributions in Eq.~(\ref{eq:Ds2317-decay}) \color{black}  are calculated through the quark rearrange mechanism, where the isospin breaking effect is caused by the mass difference between the $D^0K^+$ and $D^+K^0$ thresholds.
Thus,  the isospin breaking effects in this calculation comes from the mass difference of $u$ and $d$ quarks.

For a process in which a single-particle state  A decays
into a two-particle state BC,  the differential decay rate can be written as
\begin{equation}
d\Gamma(A \to BC)
= 2\pi \left|M_{BC,A}\right|^2
\delta^{(4)}(p_A - p_B - p_C)\,
d^3 \vec{p}_B\, d^3 \vec{p}_C .
\end{equation}
It should be noted that in the LF model, the two-body phase-space 
factor is already incorporated into the normalization of the
continuum states and, accordingly, into the definitions of the
coupling functions $f^{SL}(E)$ and the partial transition
amplitudes such as
$\langle [D_s^+\pi^0]_{S'L'}|H_I|[D^+K^0]_{SL}\rangle$~\cite{Zhou:2019swr}.
Therefore, no additional phase-space factor is required in the
width formula, and the decay width can be written directly as
\begin{align}
	&\Gamma\big(D_{s0}^*(2317)^+ \to D_s^+\pi^0\big)\nonumber\\
	&=2\pi
	\sum_{S',L'}
	\left|
	M_{S^{'}L^{'}}\left(D_{s0}^{*}(2317)\rightarrow D_{s}\pi^{0}\right)
	\right|^2.
	\label{eq:Ds2317-width}
\end{align}

\section{The quark rearrangement mechanism}

The quark rearrangement mechanism governs two parts of this calculation. The first one is the coupling between the bare meson state and the continuous meson pair states, and the second one is the rescattering of the meson pair systems. In this calculation, both procedures are OZI-allowed and they share the same set of potential parameters.

\subsubsection{The quark pair creation model}

    The OZI-allowed coupling is assumed to be described by the QPC model~\cite{Micu:1968mk,Blundell:1995ev}, in which a quark-antiquark pair with quantum $J^{PC}=0^{++}$ is assumed to be created from the vacuum and rearranged with the one in the initial meson to form the two final meson states. 
The wave functions of mock meson states are defined as in the potential model

\begin{align}
 |A(n, { }^{2s+1}l_{j\sigma})&(\vec
 P)\rangle=\sum_{m_l,m_s}\langle lm_l, sm_s
 |j\sigma\rangle\int \mathrm{d}^3\vec{p}\,
 \psi_{nlm_l}(\vec p)\,\chi^{12}_{sm_s}\,\nonumber\\
 &\times \phi^{12}\,\omega^{12}\Big|q_1\Big(\frac
 {m_1}{m_1+m_2}\vec P+\vec p\Big)\bar q_2\Big(\frac {m_2}{m_1+m_2}\vec
 P-\vec p\Big)\Big\rangle,
\end{align}
where $\chi^{12}$, $\phi^{12}$ and $\omega^{12}$  are the spin , flavor and the color wave function, respectively. $\vec{p}_1$ ($\vec{p}_2$) and $m_1$ ($m_2$) are the momentum and mass of the quark (antiquark).   $\vec P=\vec
p_1+\vec p_2$ is the momentum of the  center of mass,
and $\vec p=\frac{m_2\vec p_1-m_1\vec p_2}{m_1+m_2}$ is the
relative momentum. $\psi_{nlm_l}$ is the wave function for the meson,
$n$ being the radial quantum number.

The transition operator $T$  of creating a $q\bar q$ pair with the vacuum quantum number is defined as

\begin{eqnarray}
	\notag
     T=	-3\gamma\sum_m\langle 1 m, 1 -m|00\rangle\int \mathrm{d}^3\vec{p_3}\mathrm{d}^3\vec{p_4}\delta^3(\vec{p_3}+\vec{p_4})\\
	\times\mathcal{Y}_1^m(\frac{\vec{p_3}-\vec{p_4}}{2})\chi_{1 -m}^{34}\phi_0^{34}\omega_0^{34}b_3^\dagger(\vec{p_3})d_4^\dagger(\vec{p_4}), 
\end{eqnarray}
describing a quark-antiquark pair generated by the $b^\dagger_3$ and
$d^\dagger_4$ creation operators from the vacuum.
$\phi_0^{34}=(u\bar u+d\bar d+s\bar s)/\sqrt{3}$ is the SU(3) flavor
wave function for the quark-antiquark pair. $\chi^{34}_{1-m}$ and $\omega^{34}_0$
are the spin wave function and the color wave function, respectively.
$\mathcal Y_1^m$ is the solid Harmonic function. $\gamma$ parametrizes the production strength of the quark-antiquark pair from the vacuum. 

By considering the rearrangement of the quark constituents of $A(q_1\bar q_2)$ and the quark pair $(q_3\bar{q}_4)$ from the vacuum to form $B(q_1\bar q_4)C(q_3\bar q_2)$, one can obtain the helicity amplitude
$\mathcal M^{ABC}$ defined as~\cite{Blundell:1995ev}
\begin{equation}
	 \left\langle BC\left|T\right|A\right\rangle=\delta^3(\vec{P_f}-\vec{P_i})\mathcal M^{\sigma_A\sigma_B\sigma_C},
	 \label{eq:MABC}
\end{equation}

and its partial-wave amplitude $\mathcal M^{SL}(P)$ is
\begin{align}
    \mathcal M^{SL}(P)&=\sum_{\sigma_B,\sigma_C,M_S,M_L}\langle L M_L S M_S|j_A\sigma_A\rangle\langle j_B\sigma_B j_C\sigma_C|S M_S\rangle\nonumber\\
    &\times\int d\Omega Y^*_{LM_L}(\Omega)M^{\sigma_A\sigma_B\sigma_C}(\vec{P}).
\end{align}
Then the coupling functions $f^{SL}_{n}$ in Eq.~(\ref{eq:Ds2317-wave-funtion})  can be obtained  as
\begin{align}
f_{n}^{SL}(E)=\rho_n^{1/2}\mathcal M^{SL}(P),
\end{align}
where $\rho_n={E_BE_CP}/{E}$ is the relativistic phase space factor of the  channel $n$ with $BC$  states and the c.m. momentum of $BC$ could be written down as $P={\lambda(E,M_B,M_C)^{1/2}}/E$ where the K\"{a}llen function $\lambda(x,y,z)=x^2+y^2+z^2-2xy-2yz-2xz$. By use of these formulae, one could obtain the $f_{0+}^{SL}$ and $f_{+0}^{SL}$ defined in Eq.~(\ref{eq:Ds2317-decay}) .

The coupling amplitude $\langle [D_s^+\pi^0]_{S'L'}|H_I|c\bar s(1^3P_0)\rangle$ is obtained via the intermediate $D_s^+\eta$  followed with isospin-violated $\pi^0-\eta$ mixing
\begin{equation}
	 \langle [D_s^+\pi^0]_{S'L'}|H_I|c\bar s(1^3P_0)\rangle=\epsilon {\rho_{D_s\pi}}^{1/2}\mathcal M_{D_s\eta}^{SL}(P'),
\end{equation}
where $P'$ is the c.m. momentum of  $D_s^+\pi^0$ system similarly. The $\mathcal M^{SL}_{D_s\eta}$ decay amplitude is OZI-allowed so it is calculated by the QPC model.
The $\pi^0-\eta$ mixing factor was estimated in the leading order of chiral perturbation theory as in Ref.~\cite{Gasser:1984gg} to be
\begin{equation}
	 \epsilon=\frac{\sqrt{3}}{4}\frac{m_u-m_d}{m_s-(m_u+m_d)/2}.
\end{equation}

\subsubsection{The quark rearrangement model}
The main idea in calculating the  $DK-D_s\pi^0$
rescattering amplitude in
Eq.~(\ref{eq:Ds2317-wave-funtion}) is to evaluate the lowest order  Born diagrams
of the interchange processes of two constituent
quarks in the Barnes-Swanson (BS) model~\cite{Barnes:1991em,Barnes:1999hs}, in which the rearrangement of the constituent quark and antiquarks are also considered. 
The merit of introducing this scheme is that it shares  the same interaction potential parameters and the wave functions of the potential model, so  no additional parameter is introduced.

\color{black}
Generally, the meson-meson scattering amplitude of $AB\rightarrow CD$ could be represented by
\begin{align}
    	 S_{fi}&=2\pi i \delta( E_f-E_i)\langle CD|H_I|AB\rangle\nonumber\\
     &=2\pi i \delta(\vec P_f-\vec P_i)\delta( E_f-E_i)\mathcal{M}^{\sigma_A\sigma_B,\sigma_C\sigma_D}.
\end{align}

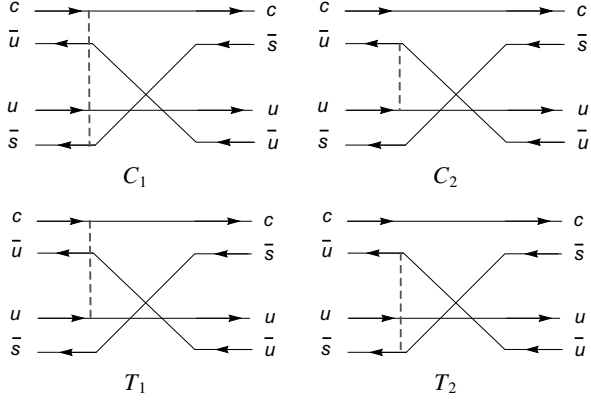
\begin{figure}[htbp]
	\centering
	
	\resizebox{0.98\columnwidth}{!}{
		\begin{tikzpicture}[
			x=1cm,
			y=1cm,
			quark/.style={line width=0.6pt},
			arr/.style={-{Latex[length=2.2mm,width=1.4mm]}},
			darr/.style={{Latex[length=2.2mm,width=1.4mm]}-},
			cut/.style={dashed,line width=1.0pt},
			labelstyle/.style={font=\large},
			subfigstyle/.style={font=\Large}
			]

			\begin{scope}[shift={(0,4)}, scale=0.82]

				\node[labelstyle] at (-0.38,2.7) {$c$};
				\node[labelstyle] at (-0.38,2.0) {$\bar{u}$};
				\node[labelstyle] at (-0.38,0.9) {$u$};
				\node[labelstyle] at (-0.38,0.2) {$\bar{s}$};

				\node[labelstyle] at (4.45,2.7) {$c$};
				\node[labelstyle] at (4.45,2.0) {$\bar{s}$};
				\node[labelstyle] at (4.45,0.9) {$u$};
				\node[labelstyle] at (4.45,0.2) {$\bar{u}$};

				\draw[quark] (0,2.7) -- (4.1,2.7);
				
				\draw[quark] (0,2.0) -- (1.1,2.0);
				\draw[quark] (2.7,2.0) -- (4.1,2.0);
				
				\draw[quark] (0,0.9) -- (4.1,0.9);
				
				\draw[quark] (0,0.2) -- (1.1,0.2);
				\draw[quark] (2.7,0.2) -- (4.1,0.2);

				\draw[arr]  (0.45,2.7) -- (0.9,2.7);
				\draw[arr]  (3.15,2.7) -- (3.65,2.7);
				
				\draw[darr] (0.45,2.0) -- (0.9,2.0);
				\draw[darr] (3.15,2.0) -- (3.65,2.0);
				
				\draw[arr]  (0.45,0.9) -- (0.9,0.9);
				\draw[arr]  (3.15,0.9) -- (3.65,0.9);
				
				\draw[darr] (0.45,0.2) -- (0.9,0.2);
				\draw[darr] (3.15,0.2) -- (3.65,0.2);

				\draw[quark] (1.1,2.0) -- (2.7,0.2);
				\draw[quark] (1.1,0.2) -- (2.7,2.0);

				\draw[cut] (1.05,0.2) -- (1.05,2.72);

				\node[subfigstyle] at (2.05,-0.48) {$C_1$};
				
			\end{scope}

			\begin{scope}[shift={(5.4,4)}, scale=0.82]

				\node[labelstyle] at (-0.38,2.7) {$c$};
				\node[labelstyle] at (-0.38,2.0) {$\bar{u}$};
				\node[labelstyle] at (-0.38,0.9) {$u$};
				\node[labelstyle] at (-0.38,0.2) {$\bar{s}$};

				\node[labelstyle] at (4.45,2.7) {$c$};
				\node[labelstyle] at (4.45,2.0) {$\bar{s}$};
				\node[labelstyle] at (4.45,0.9) {$u$};
				\node[labelstyle] at (4.45,0.2) {$\bar{u}$};

				\draw[quark] (0,2.7) -- (4.1,2.7);
				
				\draw[quark] (0,2.0) -- (1.1,2.0);
				\draw[quark] (2.7,2.0) -- (4.1,2.0);
				
				\draw[quark] (0,0.9) -- (4.1,0.9);
				
				\draw[quark] (0,0.2) -- (1.2,0.2);
				\draw[quark] (2.7,0.2) -- (4.1,0.2);

				\draw[arr]  (0.45,2.7) -- (0.9,2.7);
				\draw[arr]  (3.15,2.7) -- (3.65,2.7);
				
				\draw[darr] (0.45,2.0) -- (0.9,2.0);
				\draw[darr] (3.15,2.0) -- (3.65,2.0);
				
				\draw[arr]  (0.45,0.9) -- (0.9,0.9);
				\draw[arr]  (3.15,0.9) -- (3.65,0.9);
				
				\draw[darr] (0.45,0.2) -- (0.9,0.2);
				\draw[darr] (3.15,0.2) -- (3.65,0.2);

				\draw[quark] (1.1,2.0) -- (2.7,0.2);
				\draw[quark] (1.2,0.2) -- (2.7,2.0);

				\draw[cut] (1.05,0.9) -- (1.05,2.0);

				\node[subfigstyle] at (2.05,-0.48) {$C_2$};
				
			\end{scope}

			\begin{scope}[shift={(0,0)}, scale=0.82]

				\node[labelstyle] at (-0.38,2.7) {$c$};
				\node[labelstyle] at (-0.38,2.0) {$\bar{u}$};
				\node[labelstyle] at (-0.38,0.9) {$u$};
				\node[labelstyle] at (-0.38,0.2) {$\bar{s}$};

				\node[labelstyle] at (4.45,2.7) {$c$};
				\node[labelstyle] at (4.45,2.0) {$\bar{s}$};
				\node[labelstyle] at (4.45,0.9) {$u$};
				\node[labelstyle] at (4.45,0.2) {$\bar{u}$};

				\draw[quark] (0,2.7) -- (4.1,2.7);
				
				\draw[quark] (0,2.0) -- (1.1,2.0);
				\draw[quark] (2.7,2.0) -- (4.1,2.0);
				
				\draw[quark] (0,0.9) -- (4.1,0.9);
				
				\draw[quark] (0,0.2) -- (1.2,0.2);
				\draw[quark] (2.7,0.2) -- (4.1,0.2);

				\draw[arr]  (0.45,2.7) -- (0.9,2.7);
				\draw[arr]  (3.15,2.7) -- (3.65,2.7);
				
				\draw[darr] (0.45,2.0) -- (0.9,2.0);
				\draw[darr] (3.15,2.0) -- (3.65,2.0);
				
				\draw[arr]  (0.45,0.9) -- (0.9,0.9);
				\draw[arr]  (3.15,0.9) -- (3.65,0.9);
				
				\draw[darr] (0.45,0.2) -- (0.9,0.2);
				\draw[darr] (3.15,0.2) -- (3.65,0.2);

				\draw[quark] (1.1,2.0) -- (2.7,0.2);
				\draw[quark] (1.2,0.2) -- (2.7,2.0);

				\draw[cut] (1.05,0.2) -- (1.05,2.7);

				\node[subfigstyle] at (2.05,-0.48) {$T_1$};
				
			\end{scope}

			\begin{scope}[shift={(5.4,0)}, scale=0.82]

				\node[labelstyle] at (-0.38,2.7) {$c$};
				\node[labelstyle] at (-0.38,2.0) {$\bar{u}$};
				\node[labelstyle] at (-0.38,0.9) {$u$};
				\node[labelstyle] at (-0.38,0.2) {$\bar{s}$};

				\node[labelstyle] at (4.45,2.7) {$c$};
				\node[labelstyle] at (4.45,2.0) {$\bar{s}$};
				\node[labelstyle] at (4.45,0.9) {$u$};
				\node[labelstyle] at (4.45,0.2) {$\bar{u}$};

				\draw[quark] (0,2.7) -- (4.1,2.7);
				
				\draw[quark] (0,2.0) -- (1.1,2.0);
				\draw[quark] (2.7,2.0) -- (4.1,2.0);
				
				\draw[quark] (0,0.9) -- (4.1,0.9);
				
				\draw[quark] (0,0.2) -- (1.2,0.2);
				\draw[quark] (2.7,0.2) -- (4.1,0.2);

				\draw[arr]  (0.45,2.7) -- (0.9,2.7);
				\draw[arr]  (3.15,2.7) -- (3.65,2.7);
				
				\draw[darr] (0.45,2.0) -- (0.9,2.0);
				\draw[darr] (3.15,2.0) -- (3.65,2.0);
				
				\draw[arr]  (0.45,0.9) -- (0.9,0.9);
				\draw[arr]  (3.15,0.9) -- (3.65,0.9);
				
				\draw[darr] (0.45,0.2) -- (0.9,0.2);
				\draw[darr] (3.15,0.2) -- (3.65,0.2);

				\draw[quark] (1.1,2.0) -- (2.7,0.2);
				\draw[quark] (1.2,0.2) -- (2.7,2.0);

				\draw[cut] (1.05,0.2) -- (1.05,2.0);

				\node[subfigstyle] at (2.05,-0.48) {$T_2$};
				
			\end{scope}
			
		\end{tikzpicture}
	}
	
\caption{The four quark rearrangement diagrams of $D^0K^+\rightarrow D_s^+\pi^0$ meson-meson scatterings. The arrows represent the quark line directions.}
\label{fig:rearrange}
	
\end{figure}

The interaction is introduced through the one-gluon exchange between the constituents of $A$ and that of $B$, which is represented as the interaction potentials of Coulomb, spin-spin  and linear-confinement terms. These terms are the same as the potential model.
Four kinds of scattering diagrams, ``capture$_1$"$(C_1)$, ``capture$_2$"$(C_2)$, ``transfer$_1$"$(T_1)$, ``transfer$_2$"$(T_2)$
are considered as shown in  Fig.~\ref{fig:rearrange}, according to which
pair of the constituents are involved in the interaction~\cite{Barnes:1991em}. Thus, the helicity amplitude could be expressed as sum of contributions of four diagrams $\mathcal{M}^{\sigma_C\sigma_D,\sigma_A\sigma_B}=h_{fi}^{C_1}+h_{fi}^{C_2}+h_{fi}^{T_1}+h_{fi}^{T_2}$.

The  $T$-matrix element $h_{fi}$ of every diagram could be factorized as the
product of  flavor factor $I_{\mathrm{flavor}}$, color factor $I_{\mathrm{ color}}$, spin factor $I_{\mathrm{ spin}}$, and space factor $I_{\mathrm{space}}$
\begin{equation}
	 h_{fi}=(-1)\cdot I_{\mathrm{flavor}}I_{\mathrm{ color}} I_{\mathrm{ spin}}I_{\mathrm{space}}.
\end{equation}

 The  $(-1)$ factor comes from the odd-order permutation of fermion operators. The color factor is $(-4/9)$ for two capture diagram~($C_1$ and $C_2$) and $(4/9)$ for two transfer
diagram~($T_1$ and $T_2$). The flavor factor is obtained by the
overlap of the flavor wave functions of the initial and final states.
For the $D^{0}K^{+}\rightarrow D_{s}\pi^{0}$ process, the flavor factor is $1/\sqrt{2}$, whereas for $D^{+}K^{0}\rightarrow D_{s}\pi^{0}$
it is $-1/\sqrt{2}$, due to the flavor structure of $\pi^{0}$ meson. The spin factors for $C_1$, $C_2$, $T_1$, and $T_2$ diagrams in different interaction terms are listed in
Table~\ref{Table:spinfactor}.

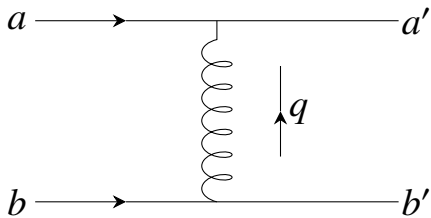
\begin{figure}[H]
	\centering
	\begin{tikzpicture}[scale=1.2, >={Stealth[length=5pt,width=5pt]}]

		\draw[->] (-2,1) -- (-1,1);
		\draw (-1,1) -- (2,1);

		\draw[->] (-2,-1) -- (-1,-1);
		\draw (-1,-1) -- (2,-1);

		\draw[decorate, decoration={coil,aspect=0.5,segment length=3mm,amplitude=2mm}]
		(0,-1) -- (0,1);

		\draw[->] (0.7,-0.5) -- (0.7,0);
		\draw (0.7,0) -- (0.7,0.5);

		\node at (-2.2,1) {\Large $a$};
		\node at (2.2,1) {\Large $a'$};
		
		\node at (-2.2,-1) {\Large $b$};
		\node at (2.2,-1) {\Large $b'$};
		
		\node at (0.9,0) {\Large $q$};
		
	\end{tikzpicture}
	\caption{Momentum redefinition in the quark-quark transition.}
	\label{Momentum}
\end{figure}

\begin{table}[H]
	\renewcommand{\arraystretch}{1.3}
	\begin{ruledtabular}
		\begin{tabular}{ccccc}
			$(S_A,S_B)\rightarrow (S_C,S_D)$ & \multicolumn{4}{c}{$(0,0)\rightarrow (0,0)$} \\
			\cline{1-1}\cline{2-5}
			& $C_1$ & $C_2$ & $T_1$ & $T_2$ \\
			\hline
			spin-spin & $-3/8$ & $-3/8$ & $3/8$ & $3/8$ \\
			Coulomb & \multicolumn{4}{c}{$1/2$} \\
			linear & \multicolumn{4}{c}{$1/2$} \\
		\end{tabular}
	\end{ruledtabular}
	\caption{\label{Table:spinfactor}Compilation of the spin factors for
		$C_1$, $C_2$, $T_1$, and $T_2$ diagram in the spin-spin hyperfine, color
		Coulomb, and linear potential terms with the total spin of two mesons being 0~\cite{Barnes:1991em}.}
\end{table}

To simplify the calculation of space matrix elements, for the quark (antiquark) transition process $q_a (\bar{q}_a) + q_b (\bar{q}_b) \to q_{a'} (\bar{q}_{a'}) + q_{b'} (\bar{q}_{b'})$,
 we redefine the momentum variables $\vec{q}=\vec{a}'-\vec{a}$,$\vec{p}=(\vec{a}'+\vec{a})/2$, as shown in Fig.~\ref{Momentum}.
Then, the space overlap factor is obtained by
\begin{equation}
	 I_{\mathrm{space}}=\frac{1}{(2\pi)^3}\int\int\mathrm{d}^3 \vec{q}\mathrm{d}^3 \vec{p}
	 \psi_C^*(\vec k_C)\psi_D^*(\vec k_D)T^{pot}_{fi}(\vec q)\psi_A(\vec k_A)\psi_B(\vec k_B),
\end{equation}
where the $\vec k_i=\frac{m_1\vec k_2-m_2\vec k_1}{m_1+m_2}$ denotes
the relative momentum of the quark-antiquark pair in the meson. 
$T^{\mathrm{pot}}_{fi}(\vec q)$ refers to the interaction potential due to color Coulomb, spin-spin hyperfine and scalar confinement interaction in the momentum representation as
\begin{equation}
	T^{\rm pot}_{fi}(\vec q) =
	\begin{cases}
		\dfrac{4 \pi \alpha_s(\vec{q})}{\vec{q}^2}, & \text{color Coulomb} \\[2mm]
		-\dfrac{8 \pi \alpha_s(\vec{q})}{3 m_i m_j}, & \text{spin-spin hyperfine} \\[1mm]
		\dfrac{6 \pi b}{\vec{q}^4}, & \text{linear confinement}
	\end{cases}
\end{equation}
where the $m_i$ and $m_j$ are the constituent quark masses of the two interacting constituents.
The overlap integrals of
wave functions in Fig.~\ref{fig:rearrange} could be written down explicitly as
\begin{widetext}
	\begin{equation}
		\begin{aligned}
		I_{\text{space}}^{C_{1}} =& \dfrac{1}{(2\pi)^{3}} \int d^3\vec{q}\, d^3\vec{p}\;
		\psi_C^{*}\left(\vec{p}+\dfrac{\vec{q}}{2}-\dfrac{1+\lambda_{C}}{2}\vec{C}\right) \,
		\psi_{D}^{*}\left(\vec{p}-\dfrac{\vec{q}}{2}-\vec{A}-\dfrac{1-\lambda_{D}}{2}\vec{C}\right)
		\\
		&\quad \times T^{\rm pot}_{fi}(\vec{q})\,
		\psi_A\left(\vec{p} -\dfrac{\vec{q}}{2} -\dfrac{1+\lambda_{A}}{2}\vec{A}\right)\,
		\psi_B\left(\vec{p} -\dfrac{\vec{q}}{2} - \dfrac{1-\lambda_{B}}{2}\vec{A}-\vec{C}\right),
		\\
		I_{\text{space}}^{C_{2}} =& \dfrac{1}{(2\pi)^{3}} \int d^3\vec{q}\, d^3\vec{p}\;
		\psi_C^{*}\left(-\vec{p}+\dfrac{\vec{q}}{2}+\vec{A}-\dfrac{1+\lambda_{C}}{2}\vec{C}\right)\, 
		\psi_{D}^{*}\left(-\vec{p}-\dfrac{\vec{q}}{2}-\dfrac{1-\lambda_{D}}{2}\vec{C}\right)
		\\
		&\quad \times T^{\rm pot}_{fi}(\vec{q})\,
		\psi_A\left(-\vec{p} +\dfrac{\vec{q}}{2} +\dfrac{1-\lambda_{A}}{2}\vec{A}\right)\,
		\psi_B\left(-\vec{p} +\dfrac{\vec{q}}{2} + \dfrac{1+\lambda_{B}}{2}\vec{A}-\vec{C}\right),
		\\	 
		\end{aligned}		 
	\end{equation}
\begin{equation}
	\begin{aligned}			
		I_{\text{space}}^{T_{1}} &= \dfrac{1}{(2\pi)^{3}} \int d^3\vec{q}\, d^3\vec{p}\;
		\psi_C^{*}\left(\vec{p}+\dfrac{\vec{q}}{2}-\dfrac{1+\lambda_{C}}{2}\vec{C}\right) \,
		\psi_{D}^{*}\left(\vec{p}-\dfrac{\vec{q}}{2}-\vec{A}-\dfrac{1-\lambda_{D}}{2}\vec{C}\right)
		\\
		&\quad \times T^{\rm pot}_{fi}(\vec{q})\,
		\psi_A\left(\vec{p} -\dfrac{\vec{q}}{2} -\dfrac{1+\lambda_{A}}{2}\vec{A}\right)\,
		\psi_B\left(\vec{p} +\dfrac{\vec{q}}{2} - \dfrac{1-\lambda_{B}}{2}\vec{A}-\vec{C}\right),
		\\
		I_{\text{space}}^{T_{2}} &= \dfrac{1}{(2\pi)^{3}} \int d^3\vec{q}\, d^3\vec{p}\;
		\psi_C^{*}\left(-\vec{p}+\dfrac{\vec{q}}{2}+\vec{A}-\dfrac{1+\lambda_{C}}{2}\vec{C}\right) \,
		\psi_{D}^{*}\left(-\vec{p}-\dfrac{\vec{q}}{2}-\dfrac{1-\lambda_{D}}{2}\vec{C}\right)
		\\
		&\quad \times T^{\rm pot}_{fi}(\vec{q})\,
		\psi_A\left(-\vec{p} +\dfrac{\vec{q}}{2} +\dfrac{1-\lambda_{A}}{2}\vec{A}\right)\,
		\psi_B\left(-\vec{p} -\dfrac{\vec{q}}{2} + \dfrac{1+\lambda_{B}}{2}\vec{A}-\vec{C}\right).
		\\
	\end{aligned}
\end{equation}
\end{widetext}
where $\lambda_X = (m_{q_{1}}-m_{\bar{q}_{2}})/(m_{q_{1}}+m_{
\bar{q}_{2}})$
is a factor describing the relative momentum of the quark–antiquark pair in meson $X$.
$\vec{X}$ denotes the momentum of particle $X$.

If the wave functions of the mesons  are described as the simple
harmonic oscillator~(SHO) wave functions, the integrations could be
simplified and the analytical transition amplitudes of all four diagrams could be expressed explicitly in terms of confluent hypergeometric functions~\cite{Barnes:1991em,Barnes:1999hs}.
However, if the meson wave functions are more realistically represented in a
large number of harmonic oscillator wave function basis, the analytical expressions
can hardly be obtained and we can only calculate the integrations
numerically in a practical manner. In calculating the terms of the
linear confinement interaction in the momentum space, the Hadamard
regularization is used to regularize the divergent integrals to obtain
the finite parts,
\begin{equation}
	\mathcal{H}\int_{0}^{\infty}\dfrac{1}{q^{2}}f(q)dq=\lim_{\epsilon \to 0^{+}} \Bigg\{\int_{\epsilon}^{\infty}\dfrac{f(q)}{q^{2}}dq-\dfrac{f(\epsilon)}{\epsilon}\Bigg\},
\end{equation}
where $f(q)$ denotes the overlap  of initial and final wave funtions in momentum space. Generally, the partial wave scattering amplitude could be obtained by 
\begin{align}
&\mathcal{M}^{S'L',SL}=\rho_{AB}^{1/2}\rho_{CD}^{1/2}\sum_{M_SM_{S'}M_LM_{L'}\sigma_A\sigma_B\sigma_C\sigma_D}\langle j_A\sigma_A j_B\sigma_B|SM_S\rangle\nonumber\\
&\times
 \langle SM_S LM_L|j\sigma\rangle  \langle j_C\sigma_C j_D\sigma_D|S'M_{S'}\rangle \langle S'M_{S'} L'M_{L'}|j\sigma\rangle\nonumber\\
&\times\int d\Omega_k\int d\Omega_{k'}\mathcal{M}^{\sigma_C\sigma_D,\sigma_A\sigma_B}
Y_{LM_L}(\hat k)Y_{L'M_{L'}}^{*}(\hat k')
\end{align}
where $M_S(M_{S'})$ is the third-component of total spin $S(S')$ and $j$ is the total angular momentum of the system with its third component $\sigma$. These formulas serve to obtain the rescattering amplitudes as $\left\langle [D_s^+\pi^0]_{S'L'}\left|H_I\right|[D^0K^+]_{SL}\right\rangle$ in Eq.~(\ref{eq:Ds2317-decay}). In this calculation, by choosing the $\hat{k}$ along the $z$-axis, the partial wave amplitude could be simplified as 
\begin{align}
&\mathcal{M}^{S'L',SL}=2\pi\rho_{AB}^{1/2}\rho_{CD}^{1/2}\sum_{\sigma_A\sigma_B\sigma_C\sigma_D}\langle j_A\sigma_A j_B\sigma_B|SM_S\rangle\nonumber\\
&\times
  \langle j_C\sigma_C j_D\sigma_D|S'M_{S'}\rangle \int d\cos\tilde{\theta}P_L(\hat{\tilde{k}})\mathcal{M}^{\sigma_C\sigma_D,\sigma_A\sigma_B}
\end{align}
with $L=L'$ and $S=S'$ where $\tilde{\theta}$ is the scattering angle of $C,D$  states w.r.t. the $z$-axis.

\section{Numerical results and discussions}
\label{result}
The wave functions of the bare meson states in the calculation are obtained from the GI model~\cite{Godfrey:1985xj}, expanded in a large number of SHO basis. To maintain theoretical consistency, the parameters used in the quark rearrangement model are also adopted from the GI model.  The running coupling constant is parametrized as
\begin{align}
    \alpha(q^2)=0.25 e^{-q^2/(1\mathrm{GeV}^{2})}+0.15 e^{-q^2/(10\mathrm{GeV}^{2})}+0.20 e^{-q^2/(1000\mathrm{GeV}^{2})},
\end{align}
and the linear parameter $b$  and the constituent quark masses are taken to be the values in Ref.~\cite{Godfrey:1985xj}:
\begin{align}
    &b=0.18 ~{\rm GeV}^2, m_c=1.628~ {\rm GeV}, m_s=0.419~{\rm GeV},\nonumber\\
    &(m_u+m_d)/2 =  0.220~{\rm GeV}, m_d-m_u = 0.005~{\rm GeV}.
\end{align}
The thresholds of $DK$ channel in LF model are taken as the values of their physical masses to ensure correct phase space factors.
Consequently, the only free parameter in this calculation is  the dimensionless parameter $\gamma$, which characterizes the vacuum quark pair creation strength.

\subsection{Determination of the $\gamma$ parameter}

\begin{figure}
    \centering
    \includegraphics[width=1\linewidth]{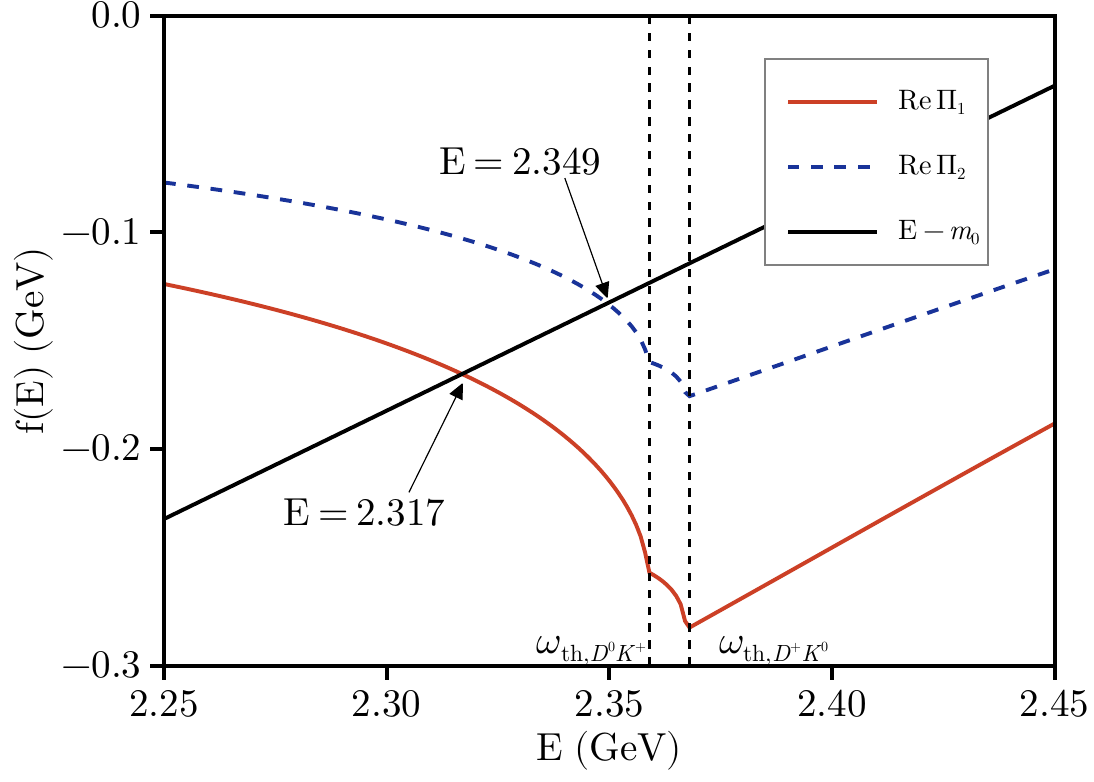}
    \caption{The intersection of the real part of $\Pi(E)$ and $E-m_0$ curves. The curves labeled $\Pi_1(E)$ and $\Pi_2(E)$ correspond to
$\gamma=8.75$ and $\gamma=6.90$, respectively. The intersection points indicate the
solutions of $E-m_0-\mathrm{Re}\,\Pi(E)=0$.}
    \label{fig:rePi}
\end{figure}

Within the LF framework, a bound-state pole emerges naturally from the $c\bar{s}(1^3P_0)-DK$ coupled channels~\cite{Zhou:2020moj}. 
This state corresponds to the intersection of the $(E-m_0)$  trajectory and the real part of the self-energy function $\Pi(E)$, as illustrated in Fig.~\ref{fig:rePi}.
If the $\gamma$ parameter is chosen to be the value commonly used in phenomenological analysis of decay width such as $\gamma=6.90$, the  bound-state pole on the first Riemann sheet is located at $z_0=2.349~{\rm GeV}$, which is only about 30 MeV above the mass of $D_{s0}^*(2317)$.  This means the bound-state pole will appear in a large range of parameter space, but the $\gamma$ parameter might not be accurate when it is determined by decay widths of other states. Here, we regard the $\gamma$ parameter as a free parameter and it is determined by requiring that the  pole position of $D_{s0}^*(2317)$ generated from the extended LF model reproduces  the physical mass of the $D_{s0}^*(2317)$. We find that 
by taking 
\begin{align}
    \gamma=8.75,
\end{align}
the  solution of the bound state pole is found at $z_0=2.317$ GeV, 
 corresponding to $D_{s0}^{*}(2317)$, located on the physical sheet below the $DK$ threshold.
 It is  acceptable  because some of the system-specific dynamical corrections are effectively incorporated into the constant $\gamma$, allowing subsequent calculations of strong decays to be more reliable and predictive.
Since the LF model provides a rigorous solution for the composition of the bound state, the fractions of the bare state ($c\bar{s}$) and the continuum components ($DK$) are naturally obtained simultaneously when finding the pole position.

\subsection{Compositeness of $D_{s0}^{*}(2317)$}

Since it is shown as a bound state, it is reasonable to define the elementariness and compositeness of the state as the Weinberg completeness relation~\cite{Weinberg:1965zz,Kinugawa:2024crb,Guo:2015daa,Aceti:2014ala,Sekihara:2014kya}
\begin{equation}
	Z=N_B^2,\,\, X_n=N_B^2\int\frac{|f^{SL}_n|^2}{(z_X-E)^2}\mathrm{d}E,
	\label{elementary}
\end{equation}
where $Z$ represents the probability of finding the discrete $c\bar{s}(1^3P_0)$ state in $D_{s0}^*(2317)$ and $X_n$ that of finding the continuous $D^0K^+$ or $D^+K^0$ state.

 Within this framework, we find that the $D_{s0}^*(2317)$ contains a bare $c\bar{s}$ component of approximately 51.1\% and a $DK$ continuum component of about 48.9\%. It is also noticed that the result is comparable with the values obtained in Ref.~\cite{Zhang:2024usz}, where the probability of the bare $c\bar{s}$ core and the $DK$ is 42\% and 58\% in a effective Lagrangian approach.
While a smaller elementariness, of about 30\%, was obtained in Ref.~\cite{Yang:2021tvc,MartinezTorres:2014kpc}, we believe that this difference primarily arises from the distinct dynamical assumptions employed, which inherently introduces systematic uncertainties that are difficult  to quantify precisely.
 In particular, the extracted elementariness and compositeness are inherently model-dependent. Their quantitative values rely heavily on how the relevant degrees of freedom are represented and  how the interactions between the continuum states and the discrete states are modeled~\cite{Guo:2015daa}.
 Therefore, the difference between the two results are not necessarily  regarded as a contradiction, but rather as an indication of the sensitivity of the extracted internal composition to the underlying dynamical framework. 
 The large $DK$ compositeness suggests that the $D_{s0}^*(2317)$ cannot be simply understood as a  bare $c\bar{s}$ state.
  Instead, it is largely generated by the strong coupling to the $DK$ continuum, and should be regarded as a dynamically generated state with a significant $DK$ molecular component, as mentioned in our previous work~\cite{Zhou:2020moj}.
 Compared with the results in Ref.~\cite{Yang:2021tvc,MartinezTorres:2014kpc,Zhang:2024usz}, there are a few discrepancy in the detailed numerical values of the components, but  a similar physical interpretation of the structure is obtained.
 This indicates that the  physical state $D_{s0}^{*}(2317)$ is a composite resonance in which the $DK$ continuum plays a crucial role.

It should be emphasized that the the determination of the components relies on the choice of the vertex functions.
Owing to the absence of a rigorous description of the coupling vertices between the bare state and the continuum state is currently unavailable,  we adopt a phenomenological description  based on the QPC model.
Here, we choose the well-accepted GI wave functions as the input, which comes from a more natural physical origin than just using  phenomenological monopole form factors.
Nonetheless, the fractions presented here should   be regarded as qualitative rather than quantitative, and the numerical values  carry inherent uncertainties. 
The elementary and compositeness provide insight into the structure of the $D_{s0}^*(2317)$, but the precise numerical values should be interpreted with caution.

 \subsection{Strong decay  width of $D_{s0}^{*}(2317)$ to $D_{s}\pi^{0}$}
Based on the obtained wave function of $D_{s0}^{*}(2317)$
 as a mixed state, it is straightforward to obtain its strong decay width to be
 \begin{equation}
 	\Gamma_{D_{s0}^{*}(2317)\rightarrow D_{s}\pi^{0}}=40.3\,\mathrm{keV},
 \end{equation}
 according to Eq.~(\ref{eq:Ds2317-width}) and other related equations.
 This small width is consistent with the observed narrow nature of the 
$D_{s0}^{*}(2317)$, since the only kinematically-allowed  decay channel is 
the $D_s\pi^0$ mode which can proceed  through 
isospin-breaking mechanisms. 
In the present calculation, the decay amplitude 
receives two main contributions: one from $\eta-\pi^0$ mixing and the other 
from the OZI-allowed $DK$ loop induced by the mass splitting between the charged and 
neutral $DK$ channels, namely $D^+K^0$ and $D^0K^+$. 

There exist already several calculations for the strong decays of the $D_{s0}^{*}(2317)$ and various
approaches give quite diﬀerent results for the  decay widths even when based
on the same structure assumption, ranging from a few keV to more than two hundred keV, as listed in Table~\ref{decaywidth}, appearing strong model dependency. It is found that in most calculations with the $D_{s0}^{*}(2317)$ regarded as a pure $c\bar{s}$ state,  its strong decay widths are usually several tens keV~~\cite{Lu:2006ry,Bardeen:2003kt,Godfrey:2003kg,Wei:2005ag,Fajfer:2015zma,Colangelo:2003vg}, while it is about one hundren keV  for the assumption of pure  $DK$
molecular states~\cite{Faessler:2007gv,Yue:2025wcl,Fu:2021wde,Su:2025aiz}.

\begin{table}[h]
	\renewcommand{\arraystretch}{1.3}
	\begin{ruledtabular}
		\begin{tabular}{ccc}
			Reference & Structure & $\Gamma(D_{s0}^{*}\rightarrow D_{s}\pi^{0})$\\
			\hline

			Ref.~\cite{Lu:2006ry} & $c\bar{s}$  &  $32$ \\
			Ref.~\cite{Bardeen:2003kt}& $c\bar{s}$  & $21.5$\\
			Ref.~\cite{Godfrey:2003kg}& $c\bar{s}$  & $\sim 10$ \\
			Ref.~\cite{Wei:2005ag} & $c\bar{s}$  &   $34\sim44$ \\
			Ref.~\cite{ Fajfer:2015zma}& $c\bar{s}$  & $2.4\sim 4.7$ \\
			Ref.~\cite{Colangelo:2003vg} & $c\bar{s}$  &  $7\pm1$  \\

			Ref.~\cite{Ishida:2003gu}& $c\bar{s}$  &  $150\pm 70$ \\
			Ref.~\cite{Faessler:2007gv}& $DK$  &  $79.3\pm 32.6$  \\
			Ref.~\cite{Yue:2025wcl}& $DK$  &    $63.0 \sim 209$\\
			Ref.~\cite{Fu:2021wde}& $DK$  &    $  132\pm 7 $\\
			Ref.~\cite{Zhou:2025rpb}& $DK$  &    $  10.3_{-2.9}^{+3.0}  $\\
			Ref.~\cite{Su:2025aiz}& $DK$  &  $140\pm10$  \\
			Ref.~\cite{Nielsen:2005zr}  & tetraquark  &  $6\pm2$    \\
			This work  & $c\bar{s}+DK$  &    $40.3^{+9.3}_{-4.9}$\\
		\end{tabular}
	\end{ruledtabular}
	\caption{\label{decaywidth} Strong decay widths of $D_{s0}^{*}(2317)\rightarrow D_{s}\pi^{0}$ in different theoretical works. The unit is keV.}
\end{table}

It should be noted that if we treat the $D_{s0}^{*}(2317)$ as  a pure compact $c\bar{s}$ state in this calculation, the width obtained is only 6 keV, which is in agreement with the result of Ref.~\cite{Godfrey:2003kg} since the wave function of the bare $c\bar{s}$ state in our calculation is obtained from the GI model. 
The non-negligible contribution from the $DK$ loop indicates that the nearby 
$DK$ continuum plays an important role in the isospin-violating transition.

Among many attempts to understand the mysterious $D_{s0}^{*}(2317)$ state, the distinct characteristic of our work is that the $D_{s0}^{*}(2317)$ emerges naturally as a dynamically generated state of the coupling between the $c\bar{s}$ core and the $DK$ continuum, and no additional parameters are introduced in this framework, which guarantees the genuine predictive power of the model. 
Its strong decay width is determined self-consistently within the same framework, without any further adjustment. In principle, such an almost-parameter-free and unified treatment could be extended to other exotic states as mentioned in Ref.~\cite{Zhou:2020moj}.

\subsection{Estimation of uncertainty}

The systematic uncertainty of this analysis could be estimated in three aspects of the theoretical model building. 

\begin{enumerate}
    \item The GI model and the BS model

In this calculation, the GI model is used to determine the bare $c\bar{s}$ spectrum and the spatial  wave functions entering the coupling vertices, whereas the   BS model is used to describe the $DK\to D_s\pi^0$ transition   through quark interchange. In principle,  the systematic uncertainties of these models enters our determination of  the absolute value of the strong decay width, since the result depends   on the adopted wave functions and transition amplitudes. However, the systematical uncertainties are difficult to evaluate. The GI model have proved highly successful in predicting the mass spectrum of  plenty of the normal states, so the statistical uncertainties of GI model parameters are  expected to be very small. However, the systematic uncertainty of the GI model and the BS model remains, and we expect these systematical uncertainties could be effectively absorbed into the variation of the $\gamma$  parameter as discussed in the following.

    \item $\gamma$ parameter

   The parameter  $\gamma$ is the sole free parameter controlling the coupling between the bare $c\bar{s}$ state and the nearby $DK$ continuum.
In this work it is determined by reproducing the physical mass of
$D_{s0}^{*}(2317)$. 
Consequently, the predicted width serves as a nontrivial consequence of the resulting mixed wave function.

In the present calculation, the possible model dependence of the
bare-continuum coupling vertex is largely encoded in the overall strength
parameter $\gamma$. When the QPC model is applied directly  to estimate the open-flavor decay width of conventional hadronic state,  comparing the experimental values with the theoretical estimates typically reveals an uncertainty of approximately 10\%~\cite{Ackleh:1996yt,Blundell:1995ev}.  Although  we find  that  the physical
mass of $D_{s0}^{*}(2317)$ is reproduced when $\gamma=8.75$, we assume that a similar uncertainty remains when using the QPC model. This variation can also  partly absorb the systematic uncertainty 
associated with the wave functions. 

To estimate this uncertainty, we vary the pair-creation strength
$\gamma$ by $\pm 10\%$ around its fitted value. We find that the
elementariness of $D_{s0}^{*}(2317)$ changes only mildly, remaining within the range
of $47.4\%$--$53.2\%$. 
This range reflects only the sensitivity of our results to the variation of the model parameter $\gamma$, since the systematic uncertainty associated with the underlying phenomenological models cannot be quantified precisely at present.
Moreover, as discussed above, the computed elementariness and compositeness are inevitably model dependent to some extent, as their quantitative values depend on the degrees of freedom and interactions explicitly included in the model.
Therefore, this range does not represent  a complete estimate of the theoretical uncertainty, but rather as an indication of the stability of the extracted composition against the variation of $\gamma$ within the present framework.
 Under this assumption, we find that the strong decay width of $D_{s0}(2317)$ is about
$40.3^{+9.3}_{-4.9}$ keV.
    \item  $\pi^0-\eta$ mixing

    The uncertainty from the $\pi^0$--$\eta$ mixing is relatively small. The
mixing parameter is an isospin-breaking quantity and is usually of the order
of $10^{-2}$. In the present calculation, we take $\epsilon_{\pi^0\eta}=0.01$,
given by chiral perturbation theory.
This value lies within the commonly used range,
$\epsilon_{\pi^0\eta}\simeq 0.009$--$0.012$, and therefore the uncertainty
associated with this input is expected to be limited.
\end{enumerate}

\section{summary}
In this work, we have studied the  compositeness of $D_{s0}^{*}(2317)$ and its strong decay width to $D_{s}\pi^{0}$ within the extended LF model combined with the quark rearrangement model.
Firstly, using the bare spectrum and wave functions as input, a bound-state pole corresponding to $D_{s0}^{*}(2317)$ is reproduced  on the physical sheet.
By matching the physical mass of $D_{s0}^{*}(2317)$ given by experiments, the sole parameter $\gamma$ is fixed.
Within this framework, the elementariness and compositeness of the bound state can be calculated by assuming that its total wave function consists primarily of a bare $c\bar{s}$ state and $DK$ continuum components.
Finally, the decay width can be computed by the QPC model and the quark rearrangement model.
The OZI-suppressed process $c\bar{s}\rightarrow D_{s}\pi^{0}$ is studied by $c\bar{s}\rightarrow D_{s}\eta$  in addition to the $ \eta-\pi^{0}$ mixing mechanism in QPC model.
The transition amplitude of $DK$ to $ D_{s}\pi^{0}$ is calculated with the quark rearrangement model and its isospin violation effect is introduced by the mass difference between the charge and neutral $D$, $K$ mesons.

The elementariness and compositeness, namely as the fractions of the bare $c\bar{s}$ and the $DK$ continuum state, are calculated as about 51.1\% and 48.9\%, respectively.
The strong decay width of  $D_{s0}^{*}(2317)\rightarrow D_{s}\pi^{0}$  is found to be  about $40.3$ keV.

The potential systematic uncertainties associated with the phenomenological potential model and the coupled-channel content of our framework are partially reflected by varying the pair-creation strength $\gamma$. 
Importantly, the quark-rearrangement calculation is fully consistent with the underlying GI model and requires no additional free parameters.
With $\gamma$ fixed by the physical mass of the $D_{s0}^{*}(2317)$, the resulting composition and isospin-violating decay width are obtained without further parameter tuning. The present approach thus provides a self-consistent and predictive framework for investigating the internal structure and isospin-violating decay of the $D_{s0}^{*}(2317)$.

\color{black}
\section*{Acknowledgement} This work is supported by China National Natural Science Foundation under contract No.12375132 and  11975075.

\bibliographystyle{apsrev4-2}
\bibliography{refof2317}

\end{document}